\documentclass{article}
\usepackage[final]{neurips_2022_ml4ps}

\usepackage[utf8]{inputenc} 
\usepackage[T1]{fontenc}    
\usepackage{hyperref}       
\usepackage{url}            
\usepackage{booktabs}       
\usepackage{amsfonts}       
\usepackage{nicefrac}       
\usepackage{microtype}      
\usepackage{xcolor}         
\usepackage{wrapfig}

\usepackage{siunitx} 
            
\usepackage{multicol}
\usepackage{changepage}
\usepackage{tabularx}
\newcolumntype{C}[1]{>{\centering\arraybackslash}p{#1}}
\usepackage{multirow,bigdelim}
\usepackage{subfig} 
\usepackage{graphicx}
\usepackage{pdfpages}

\def\ocpatches{\texttt{OC-Patches}}
\def\ocpatch{\texttt{OC-Patch}}

\usepackage{amsmath}

\DeclareMathOperator*{\argmin}{arg\,min}

\title{Insight into cloud processes from unsupervised classification with a rotation-invariant autoencoder}

\author{%
    Takuya Kurihana\thanks{Globus Labs: \url{https://labs.globus.org/}} \\
    Dept. of Computer Science\\
    University of Chicago\\
    \texttt{tkurihana@uchicago.edu} \\
\And
    James Franke \\
    Dept. of the Geophysical Sciences\\
    University of Chicago\\
    \texttt{jfranke@uchicago.edu} \\
\And
    Ian T. Foster \\
   University of Chicago \& \\
    Argonne National Laboratory \\
    \texttt{foster@uchicago.edu} \\
\And
    Ziwei Wang \\
    Dept. of the Geophysical Sciences\\
    University of Chicago\\
    \texttt{zwwang@uchicago.edu} \\
\And
    Elisabeth J. Moyer \\
    Dept. of the Geophysical Sciences\\
    University of Chicago\\
    \texttt{moyer@uchicago.edu} \\
}

\begin{document}

\maketitle

\begin{abstract}
Clouds play a critical role in the Earth’s energy budget and their potential changes are one of the largest uncertainties in future climate projections. However, the use of satellite observations to understand cloud feedbacks in a warming climate has been hampered by the simplicity of existing cloud classification schemes, which are based on single-pixel cloud properties rather than utilizing spatial structures and textures. Recent advances in computer vision enable the grouping of different patterns of images without using human-predefined labels, providing a novel means of automated cloud classification. This unsupervised learning approach allows discovery of unknown climate-relevant cloud patterns, and the automated processing of large datasets.
We describe here the use of such methods to generate a new AI-driven Cloud Classification Atlas (AICCA), which leverages 22 years and 800 terabytes of MODIS satellite observations over the global ocean. We use a rotation-invariant cloud clustering (RICC) method to classify those observations into 42 AI-generated cloud class labels at $\sim$100 km spatial resolution. As a case study, we use AICCA to examine a recent finding of decreasing cloudiness in a critical part of the subtropical stratocumulus deck, and show that the change is accompanied by strong trends in cloud classes.

\end{abstract}

\label{sec:abst}

\section{Motivation}

Stratocumulus---decks of low clouds---are critical for the climate and represent the single biggest source of uncertainty in climate response to anthropogenic forcing \citep{zelinka_causes_2020}.
The marine stratocumulus decks off the west coasts of North and South America and Africa in the sub-tropics cover only $\sim$6\% of the Earth's surface but reflect a disproportionate amount of solar radiation, keeping the planet cool. Recent studies based on relatively small-scale numerical simulations suggest that these decks are vulnerable to increases in CO\textsubscript{2} and may break up after some threshold is reached, causing an additional 8K of global  warming \citep{schneider_possible_2019,schneider_solar_2020}.
However, the global models used in most climate studies are too coarse in resolution to capture the behavior of subtropical stratocumulus, or indeed marine low clouds in general.
Current computing power does not allow simulating global clouds at the 10m resolution needed to realistically capture their behavior \citep{morrison_confronting_2020}, and will not be able to do so until at least 2060, even if computer power continues to grow at historical rates  \citep{schneider_climate_2017}.
Understanding the climate risk posed by stratocumulus transitions requires making use of satellite observations instead.

By now, satellite instruments have provided several decades of high-resolution multispectral images of the Earth. These datasets, now totaling many petabytes (PBs), should provide unprecedented opportunity to understand cloud processes.
However, satellite imagery is underemployed in cloud studies because climate scientists have had no practical means of  analyzing their spatial-temporal patterns. Traditional cloud classifications~\citep{rossow1999advances, wmo2017} have therefore been based only on mean physical properties over some area, ignoring cloud morphologies and textures.

The difficulty of extracting insight from complex patterns in PB-scale datasets motivates the application of artificial intelligence (AI) algorithms to define and label cloud types that are relevant to climate research.
However, supervised learning approaches~\citep[e.g.,][]{Lee1990ANN, Rasp2019CombiningCA, Zantedeschi2019CumuloAD,Zhang2018CloudNetGC}
which rely on human-labeled training datasets, are not useful for research where the relevant classes themselves are not known. Unsupervised learning methods, on the other hand, can  discover unknown cloud types relevant to climate change research, with definitions based on spatial distributions as well as means~\citep{kurihana2019cloud, kurihana2021data}. 
In this work, we leverage the unsupervised deep learning-based, rotation-invariant cloud clustering (RICC) approach of {our previous studies}~\citep{kurihana2019cloud, kurihana2021data} to create an atlas of cloud classes, and show that these classes can help in understanding observed trends in marine clouds. 
 





\label{sec:intro}

\section{Unsupervised classification of marine clouds in MODIS images} 

We use an unsupervised cloud clustering approach to reduce the dimensionality of more than two decades of multispectral satellite imagery from the Moderate Resolution Imaging Spectroradiometer (MODIS) instruments. 


\subsection{MODIS dataset}\label{sec:data:MODIS}

The MODIS instruments on the Terra and Aqua satellites have been capturing images of the earth since 2000 and 2002 in 36 ``moderate'' resolution spectral bands, from visible to thermal range.
Each instrument captures a $\sim$\num{2330}~km $\times$ \num{2030}~km \emph{swath} every five minutes, covering much of the globe at a daily cadence. The total dataset of raw spectral imagery is now nearly 1~PB.
We use selected fields 
from the MODIS Level 1B calibrated radiance product (\textbf{MOD02}), Level 1 geolocation fields (\textbf{MOD03}), and, for validation, Level 2 derived cloud properties (\textbf{MOD06}), all at 1~km resolution. 
We then extract $\sim$331M non-overlapping ocean-cloud \emph{patches} as the smaller geographical image unit, each of 128 pixels $\times$ 128 pixels ($\sim$100 km $\times$ 100km). We use only 6 of the 36 bands, selecting those most useful for diagnosing cloud optical properties, cloud phase, and cloud height. 
To avoid the complications of land reflectance, we further restrict sampling to ocean-only by eliminating any patches that include non-ocean pixels, using MOD06 land/water indicator. 
We also eliminate those with less than 30\% cloud pixels, as indicated by the MOD06 cloud mask, yielding a set, \ocpatches{}, of \num{146}M ocean-cloud patches.

\renewcommand{\arraystretch}{1.2}

\begin{figure}
    \vspace{-2ex}
    \begin{minipage}{.38\columnwidth}
        \subfloat[Distributions of cluster properties.\label{fig:cotctpA}]{
        \includegraphics[width=1.0\columnwidth]{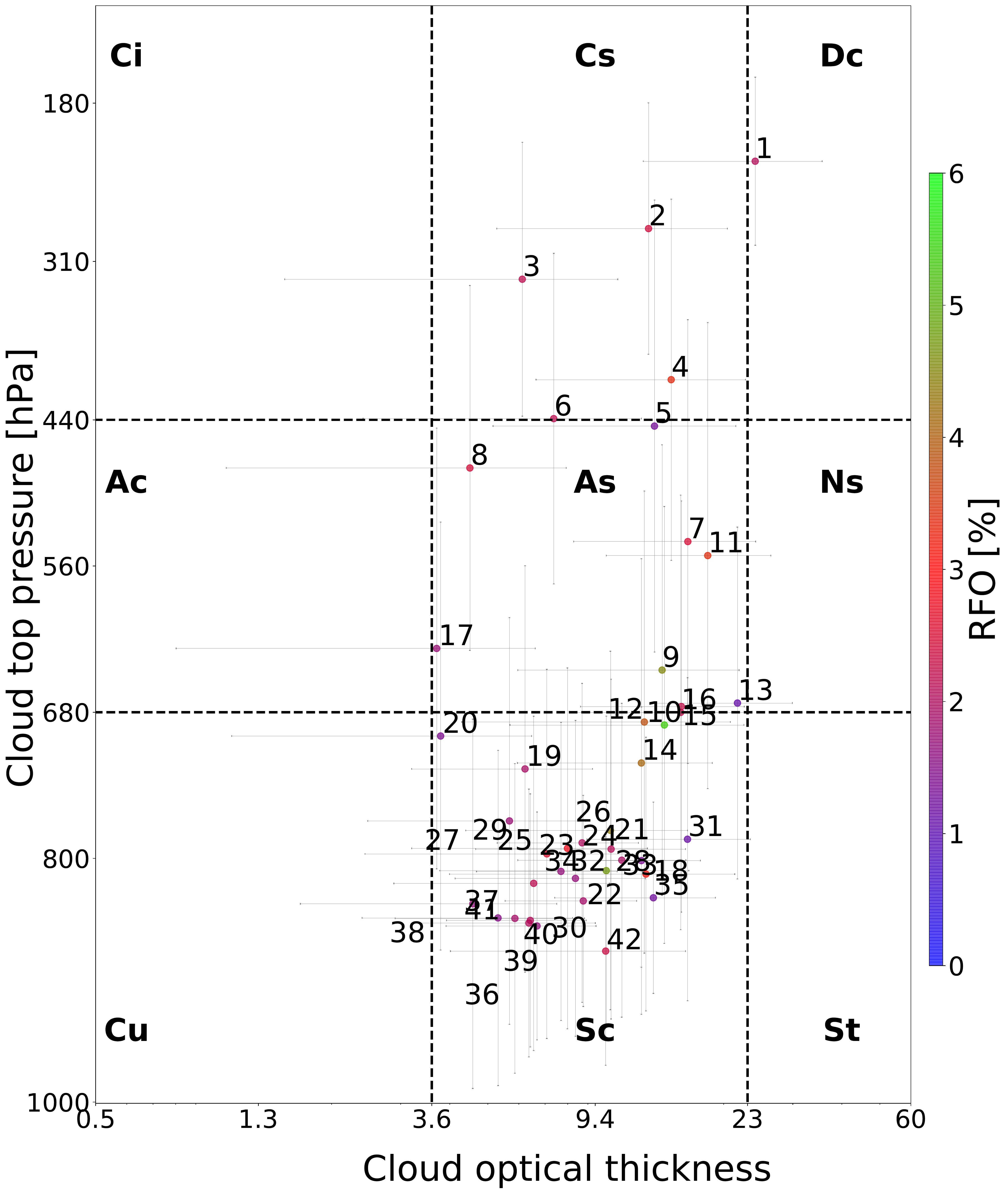}
        }
    \end{minipage}
    \begin{minipage}{.30\columnwidth}
        \subfloat[Example patches from \#3.\label{fig:c25}]{
        \includegraphics[width=.99\columnwidth]{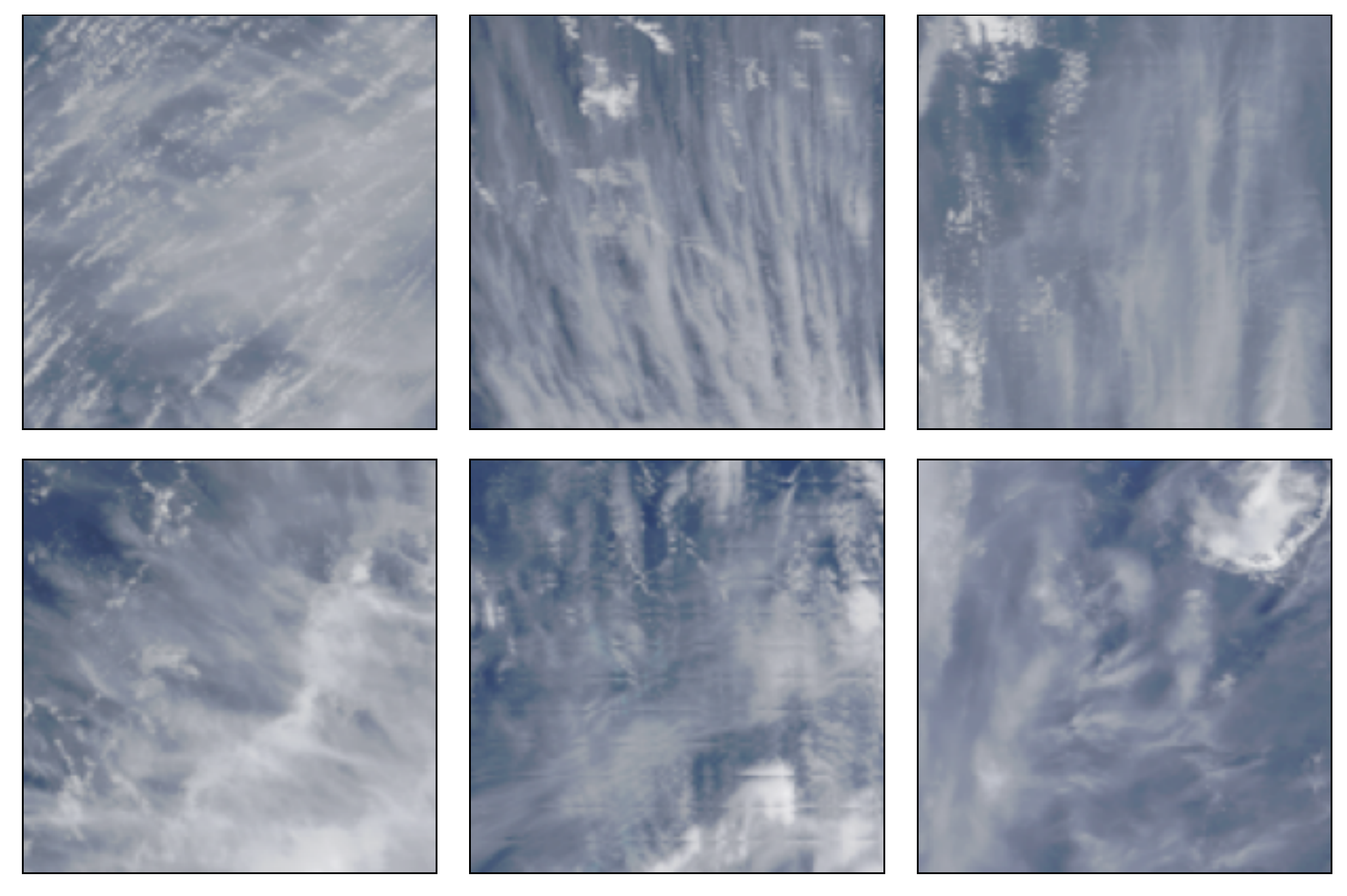}
        }        
    \subfloat[Example patches from \#35.\label{fig:c35}]{
        \includegraphics[width=.99\columnwidth]{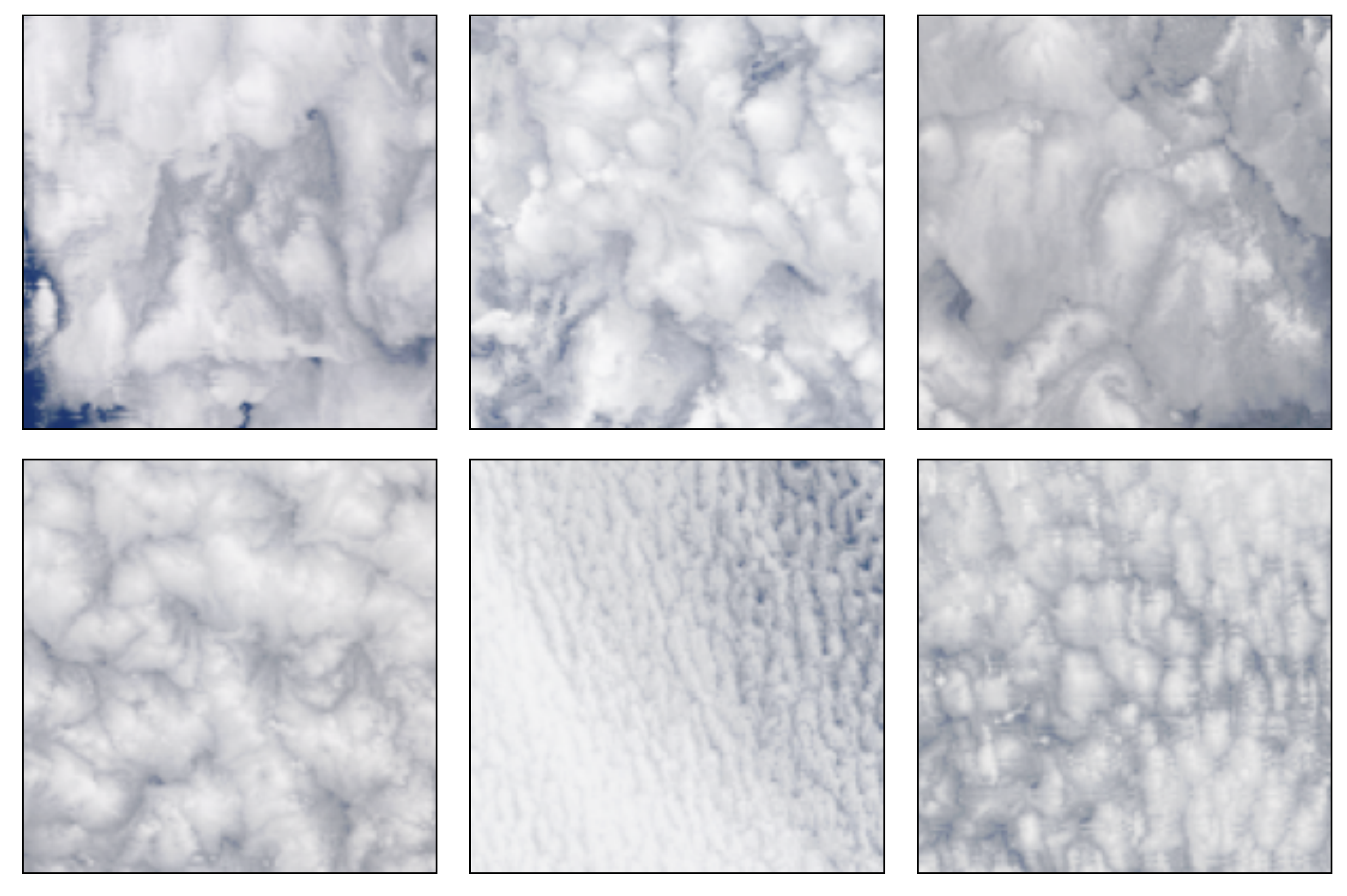}
        }    
    \vspace{0.1mm}
        \subfloat[Spatial RFO of \#3.\label{fig:c3space}]{
        \includegraphics[width=.99\columnwidth]{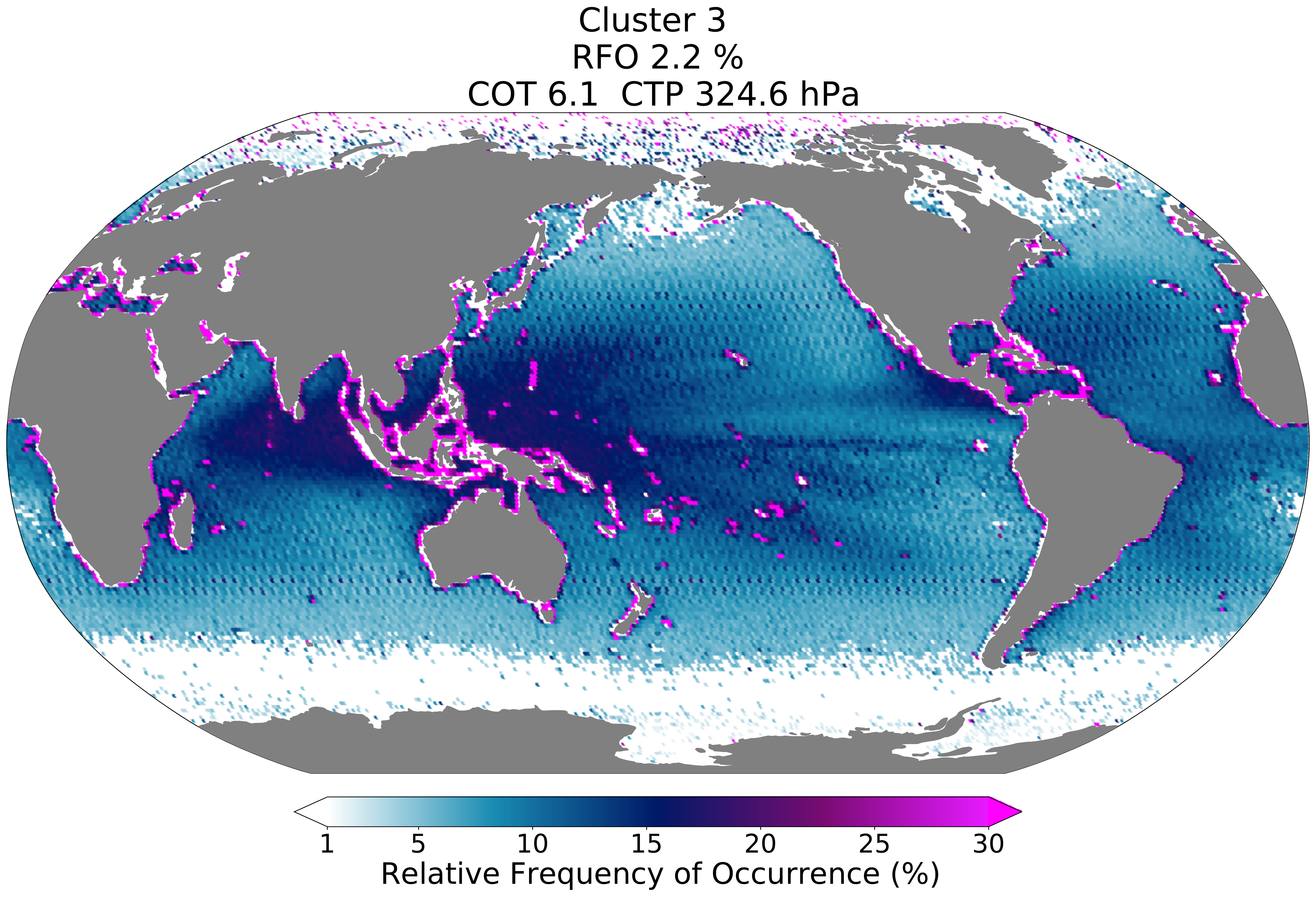}
        }
        \subfloat[Spatial RFO of \#35.\label{fig:c35space}]{
        \includegraphics[width=.99\columnwidth]{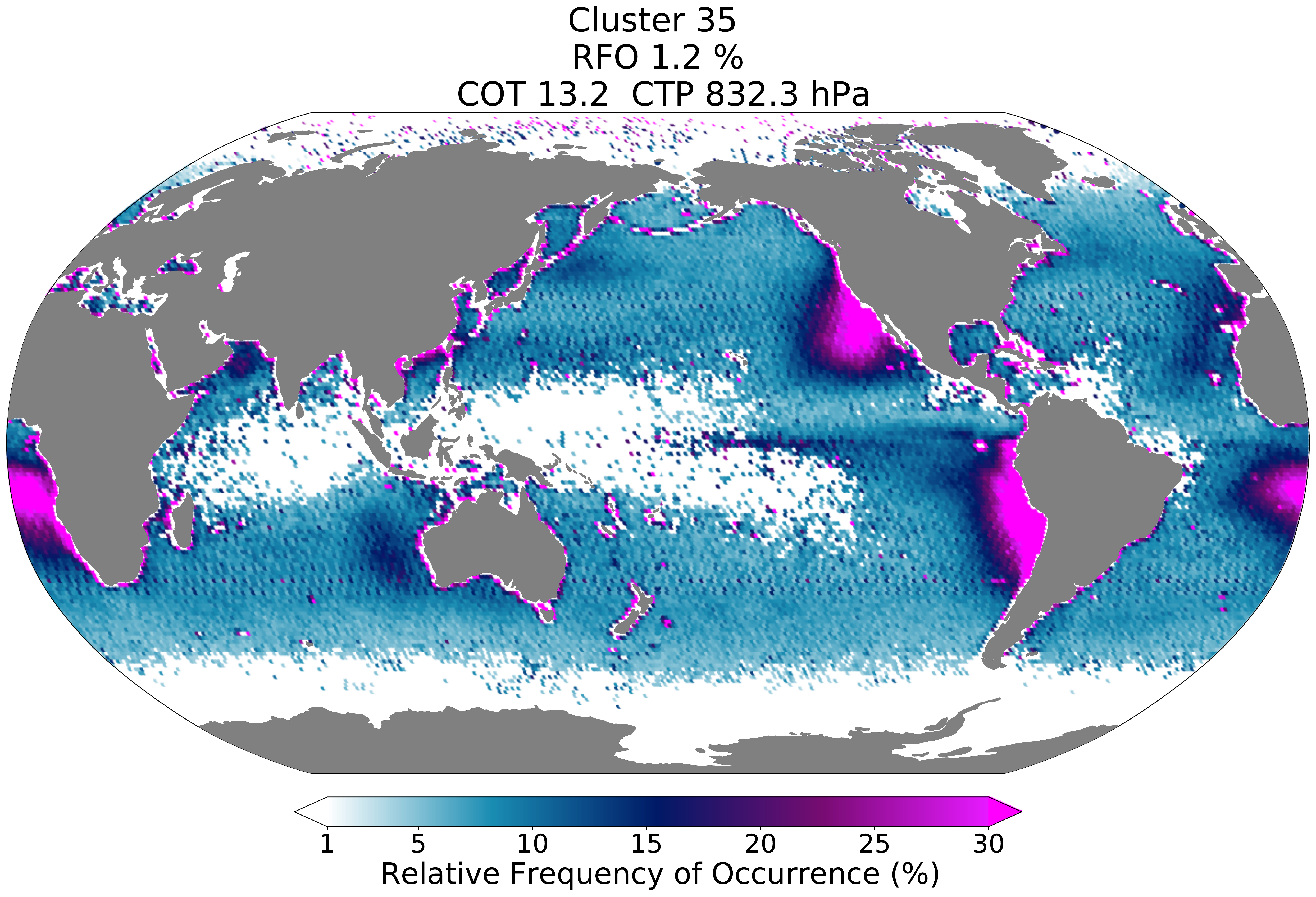}
        }
    \end{minipage}
\caption{
Demonstration of physical reasonableness of of AICCA\textsubscript{42} classes.  (a) Distributions of  AICCA\textsubscript{42} classes in cloud optical thickness (COT) -- cloud top pressure (CTP) space. Dots show mean values for each cluster and error bars show the standard deviation of properties.
For comparison, dashed lines divide the nine regions corresponding to the International Satellite Cloud Climatology Project (ISCCP) cloud classes~\citep{rossow1999advances, wmo2017}. Note that AICCA\textsubscript{42} allows far richer differentiation of cloud types, especially stratocumulus (``Sc'').  Panels (b--e) show the texture and distribution of example classes \#3 (a type of cirrostratus) and \#35 (the dominant class in the stratocumulus decks off the N.\ and S.\ American W.\ coasts).
(b--c) examples of MODIS true color images~\citep{gumley2003creating} for representative patches from each example class. (d--e) all-season geographic distribution of each example class.
}\label{fig:physicalregimes}
\vspace{-2ex}
\end{figure}

\subsection{Rotation-invariant cloud clustering: RICC}\label{sec:ricc}


The goal of the RICC method that we employ in this work~\citep{kurihana2021data} is to generate latent representations from an autoencoder that can assign cloud types to images regardless of orientation when applying hierarchical agglomerative clustering (HAC). Rotation invariance is needed because any particular physically driven cloud pattern can occur in different orientations.
A common $l^2$ loss function is insufficient to achieve this goal because it can generate different representations for an image $x$ and a rotated version of the same image $\mathcal{R}(x)$ (where ${\cal R}$ is a set of scalar rotation operators), which may then be placed in different clusters. 
The RI autoencoder is inspired by a shifted transform invariant autoencoder~\citep{Matsuo2017TransformIA}. It optimizes $L_{\mathrm{RI}}$ (see Equation~\ref{eq:ri}), which combines both a rotation-invariant loss $L_{\mathrm{inv}}$ (which learns the rotation-invariance, to map different orientations of identical input images into an identical representation) and a restoration loss $L_{\mathrm{res}}(\theta)$ (which learns the important spatial structures) to restore spatial patterns in inputs via RI autoencoder.
The RI loss is formulated as:
\begin{equation}\label{eq:ri}
    L_{\text{RI}}(\theta) =  \underbrace{\frac{1}{N}\sum_{x \in S}\sum_{R\in {\cal R}} {|| D_{\theta}(E_{\theta}(x)) - D_{\theta}(E_{\theta}(R(x)))||^2_2}}_{L_{\mathrm{inv}}(\theta)} + \underbrace{\sum_{x \in S} \displaystyle \min_{R \in {\cal R}} {|| R(x) - D_{\theta}(E_{\theta}(x)) ||^2_2}}_{L_{\mathrm{res}}(\theta)} ,
\end{equation}
where $E$ and $D$ represent the encoder and decoder, respectively; $S$ are training images; and the scalars $\lambda_{\text{inv}}$ and $\lambda_{\text{res}}$ are the weights of the respective loss terms. (We use $ \left(\lambda_{ \text{inv}},\lambda_{ \text{res}} \right) = \left( 32, 80\right) $.)
Optimizing the values of $\theta$ produces a latent representation that preserves the spatial structure in $x$ and transforms $x$ and a set of ${\cal R}(x)$ to a canonical orientation. 
Kurihana et al.\ train the autoencoder at learning rate 0.01 for 100 epochs with 32 NVIDIA V100 GPUs.
{Our} code is available in Github: \url{https://github.com/RDCEP/clouds}.

\subsection{AICCA: An AI-driven Cloud Classification Atlas}\label{sec:aicca2}
We construct AICCA by assigning a cluster label to each of the 146M patches in \ocpatches{}. For each \ocpatch{} $x_i$, we calculate the cluster label assignment $c_{k,i}$ by identifying
the cluster centroid $\mu_k$ with the smallest  Euclidean distance to its latent representation, $z(x_i)$:
\begin{equation}\label{eq:inference}
    c_{k,i} =  \argmin_{k=\{1,\cdots, k^{\ast}\} }{|| z(x_i)  - \mu_k ||_2 } ,
\end{equation}
where $\mu_k$ is a set of $k^{\ast}$ cluster centroids by applying a trained RI autoencoder with Equation~\ref{eq:ri} on 1M patches selected at random from \ocpatches{} and applied HAC to the \num{74911} ocean-cloud patches from the year 2003.
We select $k^{\ast}$ = 42 by following a {five-criteria evaluation protocol based on quantitative and qualitative tests}
so that the dataset is formally ``AICCA\textsubscript{42}'', but for convenience we use here simply ``AICCA''{; see~\citep{kurihana2021data, kurihana2022Aicca} for further details on the protocol.}
The evaluation protocol addresses the difficulty in unsupervised learning that there is no perfect ground truth against which output can be compared. 

The resulting AICCA dataset displays strong explanatory power (\autoref{fig:physicalregimes}). It distributes classes in 
COT--CTP space in a manner of consistent with occurrence frequencies (panel a). 
The 42 AICCA classes allow far richer cloud information about stratocumulus than does the conventional 9-class scheme: 30 of these classes cover the space of the conventional ``Sc'' cloud type.
The AICCA  classes show consistent textures and physically reasonable properties (panels b--c) and physically reasonable geographic distributions (panels d--e). 
Examples shown here are tropical high-altitude clouds that would be broadly termed ``cirrostratus'' (class \#3) and low-altitude stratocumulus decks that form along the West coast of N.\ and S.\ America (class \#35).

\label{sec:aicca}

\section{Trends in subtropical stratocumulus}

For our case study, we consider whether the AICCA classes can provide insight into a recently discovered cloud trend: low cloud cover has been decreasing in the Pacific Ocean off the coast of Southern California and the Baja peninsula \cite{andersen_attribution_2022}. This finding was based on mean MODIS cloud properties alone, with no consideration of cloud patterns. Trends in AICCA classes suggest that the reduction is driven by losses in stratocumulus decks, but that other classes of sparse, optically thinner clouds actually increase in frequency.

\begin{figure}[ht!]
    \centering
    \includegraphics[width=\textwidth]{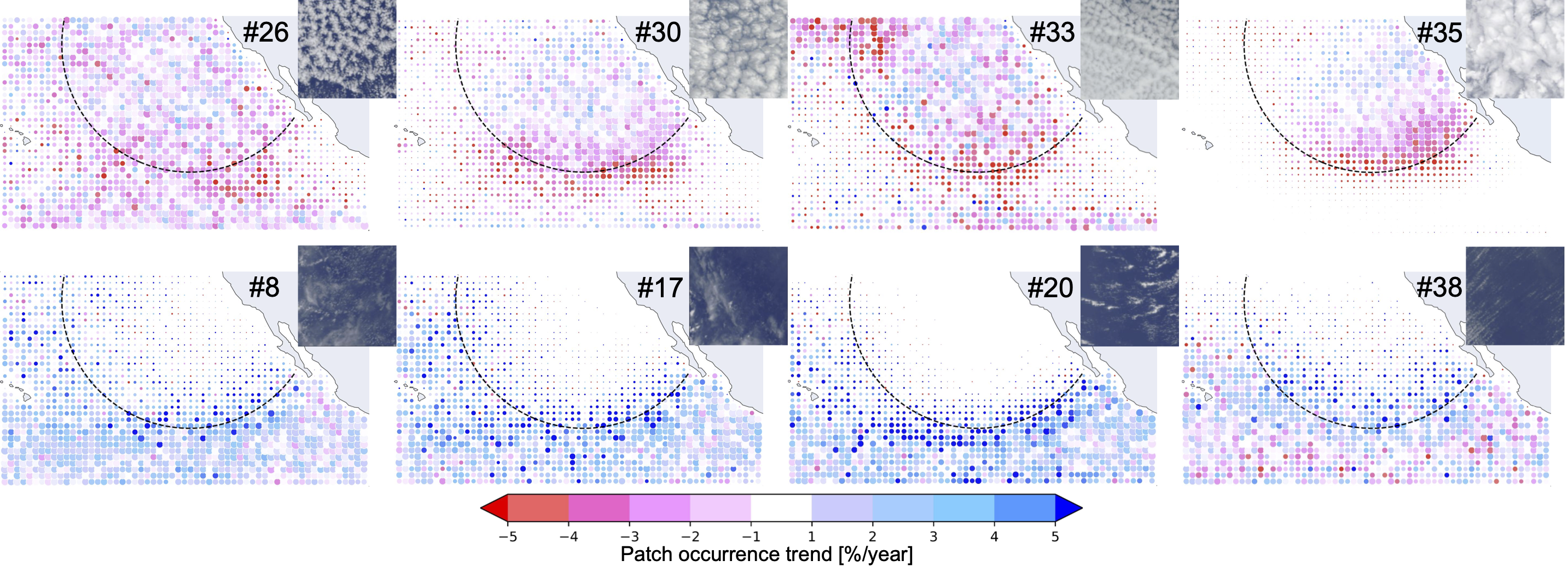}
    \caption{Trends in occurrence of selected AICCA cloud classes over the subtropical N.\ Pacific Ocean (100--160W and 5--40N). Colors show fitted linear trends over the 18 years 2003--2021, expressed in units of \% of the mean value over this time. Dot size represents mean relative frequency of occurrence within a class. 
    Dashed curve highlights the approximate edge of the deck.}
    \label{fig:trends}
\end{figure}

\autoref{fig:trends} shows the trends in 18 years (2003--2021) of MODIS observations off the S.\ California coast for eight cloud classes.  We exclude 2000--2002 because the Aqua and Terra instruments were not operating simultaneously over an entire year during that period. Closed-cell (or transitional) stratocumulus are the most dominant cloud types in this part of the Pacific, making up a quarter of all cloud occurrences (top row, \#26, 30, 33, 35, which are four of the five most predominant classes). 
All these classes decrease strongly, especially along the southern edge of the deck---by as much as 2/3 in some locations.
However, several classes of optically thinner clouds increase, especially just south of the deck region (bottom row, \#8, 17, 20, and 38). These classes represent very sparse cumulus or alto-cumulus, similar in visual texture but slightly higher in altitude.
The combined effect is that sparse classes infiltrate the stratocumulus deck along its unstable edges. 
{(The same analysis applied to ISCCP cloud classifications also shows a decrease in stratocumulus and increase in cumulus clouds, but AICCA cloud classes provide finer details on these transitions.)}
 This trend may be temperature-driven: 
 {stratocumulus decks form only when atmospheric conditions are highly stable, and warmer sea surface temperatures tend to reduce stability.}
 Sea-surface temperatures in the region from the ERA5 reanalysis \citep{hersbach_era5_2020} 
do increase by nearly 1K over this time period.

 

\begin{wrapfigure}[21]{r}{0.47\textwidth}
\vspace{-5mm}
  \begin{center}
\includegraphics[width=0.47\textwidth,trim=3mm 2mm 3mm 3mm,clip]{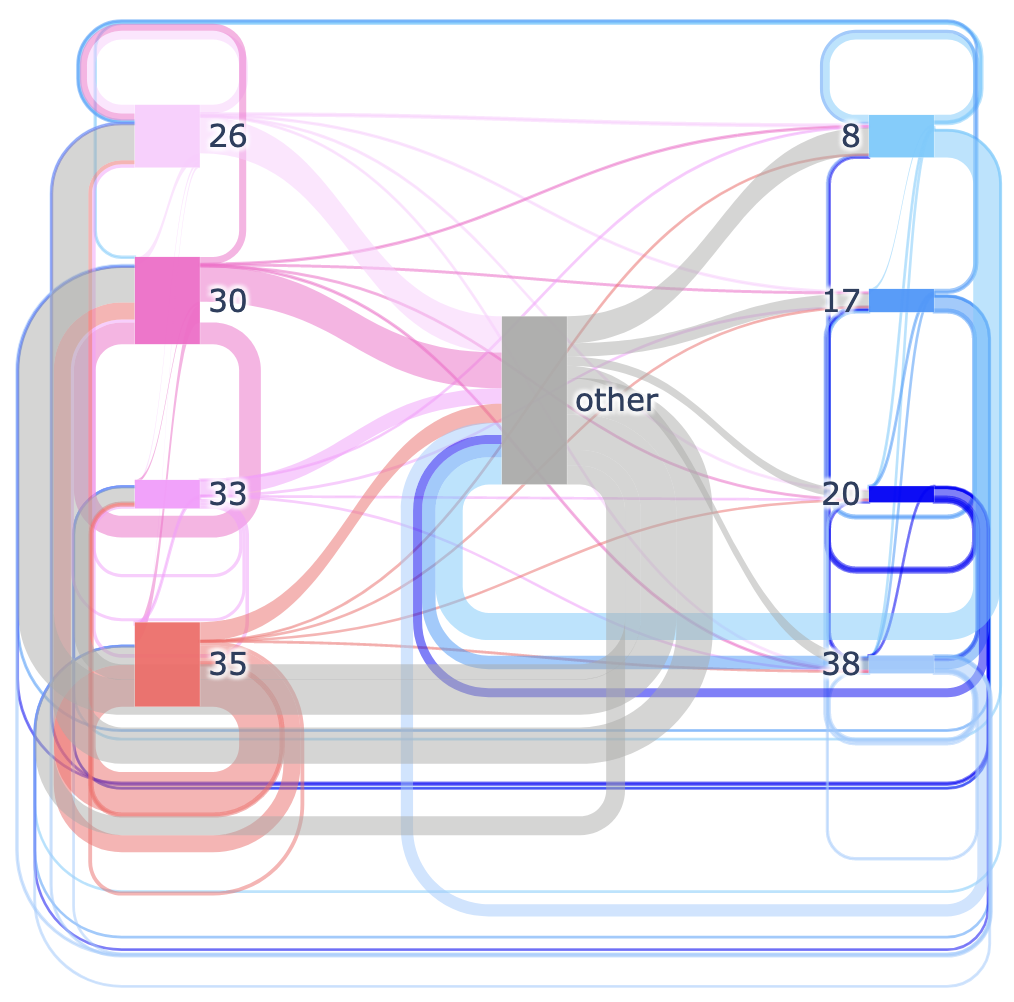}
  \end{center}
  \vspace{-4mm}
  \caption{Sankey diagram of sub-daily and daily cloud class transitions within gridcells for the 2003--2022 period for the eight classes, and region, in \autoref{fig:trends}. Box widths represent  occurrence frequency of each class, and colors are ordered by trend. Ribbon width represents the rate of transitions from one class type to another. Loops represent persistence of a class from one time period to the next.}  \label{fig:pathways}
\end{wrapfigure}

The AICCA dataset also lets us examine the temporal evolution of cloud patterns over shorter timescales. For now we consider evolution at fixed (Eulerian) locations, {ignoring advection by winds. (Stratocumulus textures are driven by a mix of local environmental control and horizontal advection, but the Eulerian perspective can provide insight at sub-daily to daily timescales.)}
\autoref{fig:pathways} shows, as a Sankey diagram, transitions from one time period to the next ($\sim$ daily) for the selected classes and region shown in \autoref{fig:trends}.
We see significant evolution within closed-cell stratocumulus.
The most frequently occurring class, \#35, is unsurprisingly also the most stable, and evolves primarily into other closed-cell forms (\#30 and \#26), which in turn transition into a wide variety of classes. Less than 10\% of transitions represent direct evolution from selected closed-cell to selected sparse cloud classes (though still more than the reverse direction).
{That is, transitions from closed cell classes to sparse cloud classes are largely modulated through a complicated network of mixed type classes.}
This result suggests that the replacement of closed-cell by sparse cloud types in \autoref{fig:trends} does not simply result from a reduction in stratocumulus lifetime.



\section{Summary}\label{sec:summary}
In summary, the unsupervised cloud clustering approach employed in this work enables reducing the dimensionality of multidecadal MODIS satellite imagery into a tractable set of physically meaningful cloud clusters. It permits the discovery of new classes based on both cloud morphology and physical properties that are unbiased by artificial assumptions and  
that capture the diversity of global cloud types.
We have shown an illustrative example of how the AICCA dataset produced via this approach can be used to track both long-term trends and short-term transitions in cloud morphology, important for understanding cloud feedbacks.
More broadly, this work shows the promise of unsupervised learning
for unlocking new applications of large satellite datasets in Earth science. 
\label{sec:results}

\newpage
\section*{Broader Impact}
The AICCA dataset that we present here, generated via the unsupervised cloud clustering approach of Kurihana et al., has both immediate uses for cloud research and broader application in Earth science generally. 

The AICCA dataset can be applied immediately to some of the most urgent questions in climate science: how clouds alter under conditions of higher CO\textsubscript{2} and warmer temperatures. The dataset is designed in particular to address questions about marine stratocumulus clouds, whose potential breakup is the largest uncertainty in climate projections. While observations will provide the most direct insight, the same approach can also be applied to analyzing the fidelity of cloud representations in Earth system models, which are producing increasingly realistic cloud distributions. Other promising lines for future analysis include the impacts aerosols or of large scale circulation pattern impacts such as ENSO.

More broadly, as the volume of data in Earth sciences explodes, unsupervised deep learning-based classification methods can find uses in many areas. Advances in measurement techniques have brought a deluge of new data, from PBs of satellite images or model output to location data from animal tracking to 3D scanning of fossil collections. Some means is needed to make these datasets tractable. Unsupervised deep learning methods can deliver these complex datasets in compact forms that allow interpretation. Finally, by facilitating access to core data, unsupervised methods can also help democratize climate research, reducing the barriers to entry in the field.

Unsupervised approaches, like any black-box interpretive tool, do offer some dangers. The proliferation of automated classifications not directly based on physical principles could negatively impact scientific fields by pulling researchers further from physical insight. As unsupervised classification grows in popularity, it will be important to produce community standards on how to validate the meaningfulness of the categories produced. Overall, however, the large potential benefits make its use worthwhile.




\label{sec:impact}

\begin{ack}
Computing resources were provided by the University of Chicago Research Computing Center and the Argonne Leadership Computing Facility (ALCF). 
This work was supported in part by the U.\ Chicago Center for Robust Decision-making on Climate and Energy Policy (RDCEP), funded by NSF grant SES-1463644 through the Decision Making Under Uncertainty program, and the U.S.\ Department of Energy under Contract DE-AC02-06CH11357.
T.K, J.A.F., and Z.W. were supported by the NSF NRT program (grant DGE-1735359).
J.A.F.\ was supported by the NSF Graduate Research Fellowship Program (grant DGE-1746045).
Codebases are available in Github: \url{https://github.com/RDCEP/clouds}.
\end{ack}

\bibliography{references}



\end{document}